\documentclass[apj]{emulateapj}
\usepackage{amsmath}
\usepackage{graphicx}
\usepackage{booktabs,threeparttable}
\usepackage{hyperref}
\usepackage{multirow}
\usepackage{ulem}
\def \na {NewA}
\shorttitle{Fantastic Striations and Where to Find Them}
\shortauthors{Chen~et al.}
\begin{document}
\title{Fantastic Striations and Where to Find Them:\\
%Formation of Dense Sub-Layers from Secondary Converging Flows in Oblique MHD Shocks and 
Origin of Magnetically Aligned Striations in Interstellar Clouds}
\author{Che-Yu Chen\altaffilmark{1}, Zhi-Yun Li\altaffilmark{1}, Patrick K. King\altaffilmark{1,2}, and Laura M. Fissel\altaffilmark{3}}
\altaffiltext{1}{Department of Astronomy, University of Virginia, Charlottesville, VA, 22904}
\altaffiltext{2}{Harvard-Smithsonian Center for Astrophysics, Cambridge, MA, 02138}
\altaffiltext{3}{National Radio Astronomy Observatory, Charlottesville, VA, 22904}

% dense, compact, sheet-like -> sub-layer (compact sheet?)
% overdense, condensed, massive -> cores, filaments

\begin{abstract}

Thin, magnetically aligned striations of relatively moderate contrast with the background are commonly observed in both atomic and molecular clouds.  They are also prominent in MHD simulations with turbulent converging shocks. The simulated striations develop within a dense, stagnated sheet in the mid plane of the post-shock region where magnetically-induced converging flows collide. 
We show analytically that the secondary flows are an inevitable consequence of the jump conditions of oblique MHD shocks. 
%We show analytically that the secondary converging flows are an inevitable consequence of the conservation of momentum and mass, as expressed by the jump conditions of oblique MHD shocks. 
They produce the stagnated, sheet-like sub-layer through a secondary shock when, roughly speaking, the Alfv{\'e}nic speed in the primary converging flows is supersonic, a condition that is relatively easy to satisfy in interstellar clouds. The dense sub-layer is naturally threaded by a strong magnetic field that lies close to the plane of the sub-layer. The substantial magnetic field makes the sheet highly anisotropic, which is the key to the striation formation. Specifically, perturbations of the primary inflow that vary spatially perpendicular to the magnetic field can easily roll up the sheet around the field lines without bending them, creating corrugations that appear as magnetically-aligned striations in column density maps. On the other hand, perturbations that vary spatially along the field lines curve the sub-layer and alter its orientation relative to the magnetic field locally, seeding special locations that become slanted overdense filaments and prestellar cores through enhanced mass accumulation along field lines.
% tilting away from the magnetic field?
In our scenario, the dense sub-layer unique to magnetized oblique shocks is the birthplace for both magnetically-aligned diffuse striations and massive star-forming structures.

%Elongated striations aligned with local magnetic field are commonly seen in molecular clouds, but the forming mechanism of these low-density sub-filaments has been under debate. We investigated the striations established in core-forming simulations with converging flows, and identified a thin, stagnated sub-layer located at the mid plane of the shock-compressed region. This dense sub-layer is created by secondary convergent flow roughly along the magnetic field in the post-shock layer, 
%%and provides a preferred environment for overdense structures (cores, filaments, striations) to form. 
%and its corrugation is spatially correlated with the striations.
%We therefore compared these simulated striations with gas structures formed in test models with corrugated central sub-layer generated by simplified velocity turbulence.
%Our results suggest that the initial velocity variation in the large-scale inflow, if parallel to the post-shock magnetic field, may be the key determining the birth site of dense filaments and cores.
%Velocity variation perpendicular to the post-shock magnetic field, on the other hand, may induce shearing force that is strong enough to twist the magnetic field and further corrugate the central sub-layer. 
%We conclude that the projections of the corrugations in the sub-layer can explain the measured features of both observed and simulated striations.

\end{abstract}

\section{Introduction}
\label{sec:intro}

Striations are highly elongated features in interstellar clouds that are often aligned with magnetic fields. One of the best examples can be found in the Riegel-Crutcher cloud of cold neutral medium (CNM), where magnetically aligned, ``hair-like" features are observed in 21cm HI absorption against the bright background toward the Galactic center region \citep{2006ApJ...652.1339M}. Such features are now known to be common in diffuse atomic medium \citep{2014ApJ...789...82C}, based on the Galactic Arecibo L-Band Feed Array HI (GALFA-HI) survey \citep{2011ApJS..194...20P}. They are also detected in the Taurus molecular cloud through $^{12}$CO observations \citep{2008ApJ...680..428G} and appear to be fairly common in molecular clouds (MCs), based on {\it Herschel} observations of dust continuum emission (\citealp{2012A&A...543L...3H,2013A&A...550A..38P}; see review in \citealp{2014prpl.conf...27A}). 
In general, striations are relatively diffuse structures with column density $\sim 2$ times of the background value \citep{2008ApJ...680..428G}, and are not necessarily associated with denser components (filaments or prestellar cores) of the clouds \citep[see e.g.~the Polaris flare,][]{2010A&A...518L.104M}.
Though visible across a range of spectroscopic channels \citep[about $2-4$~km/s in velocity difference;][]{2006ApJ...652.1339M,2008ApJ...680..428G}, striations are indeed spatially coherent structures within small velocity intervals \citep[less than $0.25$~km/s;][]{2016MNRAS.461.3918H}. More importantly, velocity channel images suggest that there is no significant velocity variation along individual striations \citep{2006ApJ...652.1339M,2016MNRAS.461.3918H}.
%M+06: 
%G+08: in the lower level emission / velocity 5-8 km/s / subthermally excited / N ~ 2e21 cm^-2 (~2x background; relatively diffuse) / parallel to B measured by optical starlight polarization / **spatially coherent within velocity intervals less than 0.25 km/s** (H+16)
%H+08: velocity 5.5-7.5 km/s
%H+16: channel images of 12CO and 13CO / evident within the core velocity interval 6.3-6.7 km/s / also have faint features within 7.38-7.72 km/s / Figure 1: no significant velocity variation within single striation! / quasi-periodic pattern/velocity oscillation / faint 13CO compared to 12CO

%- [some other attempts trying to explain striations theoretically: turbulent anisotropy (Ostriker+01, Vestuto+03, Heyer+08), MHD wave (Heyer+16, Tritsis \& Tassis 16)]
There have been a few attempts on explaining the origin of striations theoretically. One possible cause for this striped pattern is from the anisotropy of MHD turbulence at small scales \citep{1995ApJ...438..763G}. In a magnetized medium, the turbulence energy tends to have more power for wavenumbers $\hat{k} \perp \mathbf{B}$. Therefore, if the magnetic field is sufficiently strong in the clouds, the turbulent anisotropy leads to the formation of striations/sub-filaments with small separations aligned parallel to local magnetic field. Numerically, \citet{2003ApJ...590..858V} and \citet{2008ApJ...680..420H} found that in order to have significant turbulent anisotropy and prominent striation pattern, the plasma $\beta$ must satisfy $\beta \lesssim 0.2$. This value is later affirmed by \citet[][hereafter CO14, CO15]{2014ApJ...785...69C,2015ApJ...810..126C} in their core-forming simulations for when striations are clearly seen in certain models.

More recently, \cite{2016MNRAS.461.3918H} examined the striations identified in $^{12}$CO and $^{13}$CO emission maps of the Taurus molecular cloud, and proposed either the Kelvin-Helmholtz instability or magnetosonic waves as the origin of this quasi-periodic pattern of density/velocity oscillation. Similarly but independently, \cite{2016MNRAS.462.3602T} also considered these two mechanisms as the forming mechanism of striations as well as velocity gradients between streamlines parallel to the magnetic field. Based on their ideal MHD simulations, they concluded that non-linear coupling between MHD waves and inhomogeneous gas material is the most plausible scenario to create striations.

%- [our approach: starting from investigating simulated striations naturally form in star-forming regions]
In this paper, we propose a completely different scenario for the formation of the striations. We arrived at this scenario through analyzing the striations that naturally emerged in some of existing MHD simulations of star-forming clouds with inherent turbulent converging flows (\hyperlink{CO14}{CO14}, \hyperlink{CO15}{CO15}). 
%We started our study from investigating simulated striations naturally emerged in star-forming regions, which are formed under the compression of MHD shocks (\hyperlink{CO14}{CO14}, \hyperlink{CO15}{CO15}). 
By making slices through striations along different directions and plotting density profiles of these slices, we were able to view the 3-D structure of gas material around striations. We discovered the existence of a dense sub-layer located at the central plane of the post-shock region, within which prestellar cores, filaments, and striations form. 
%We concluded that the corrugations perpendicular to the magnetic field of this stagnated sub-layer is the key to striation formation.
This eventually led to our new scenario of striation formation from corrugation of the dense, stagnated sub-layer in a direction least resisted by the tension of the magnetic field threading the sheet. We believe the scenario is quite general because magnetic fields and shocks are ubiquitous in interstellar clouds, and the striation-forming dense sub-layer is an unavoidable consequence of oblique MHD shocks.  

The outline of the rest of the paper is as follows. We provide an analytic derivation for downstream gas dynamics of oblique MHD shock in Section~\ref{scf}, and demonstrate the existence of a dense, thin sub-layer in the central plane of the post-shock region.
Such a sub-layer could be the birthplace of not only relatively diffuse striations, but also overdense star-forming filaments, as illustrated by the numerical simulation described in Section~\ref{sec:simstri}. 
%The numerical simulation and the striations formed wherein are described in Section~\ref{sec:simstri}. 
General properties of striations identified in this core-forming simulation are also discussed in Section~\ref{sec:simstri}. 
In Section~\ref{sec:turb}, 
we perform simplified numerical experiments to highlight various aspects of the proposed scenario for striation formation. 
%we introduce the simplified turbulent models adopted in this study in order to examine the forming mechanism of striations,
%We illustrate and discuss in detail the features of gas structure formed under different setups of gas turbulence. 
%and deliberate features of gas structure induced under various setups of pre-shock velocity turbulence.
%and compare them to the simulated striations investigated in Section~\ref{sec:simstri}.
%The summary and conclusions are listed in Section~\ref{sec::sum}. 
Further discussions are provided in Section~\ref{sec::disc}, while the summary of our main conclusions is listed in Section~\ref{sec::sum}.

%\section{Observed Striations: Review}
%\label{sec:obsstri}

\section{Gas Dynamics in Oblique MHD Shocks}
\label{scf}

We will start with a discussion of oblique MHD shocks that are central to our investigation. They have been briefly explored in the appendix of \citet[][hereafter CO12]{2012ApJ...744..124C}, and a more detailed derivation of jump conditions is presented in \hyperlink{CO14}{CO14}. Here we focus on studying the post-shock gas dynamics and its relation with the magnetic field.

\subsection{Secondary Convergent Flows}

\begin{figure}
\begin{center}
\includegraphics[width=\columnwidth]{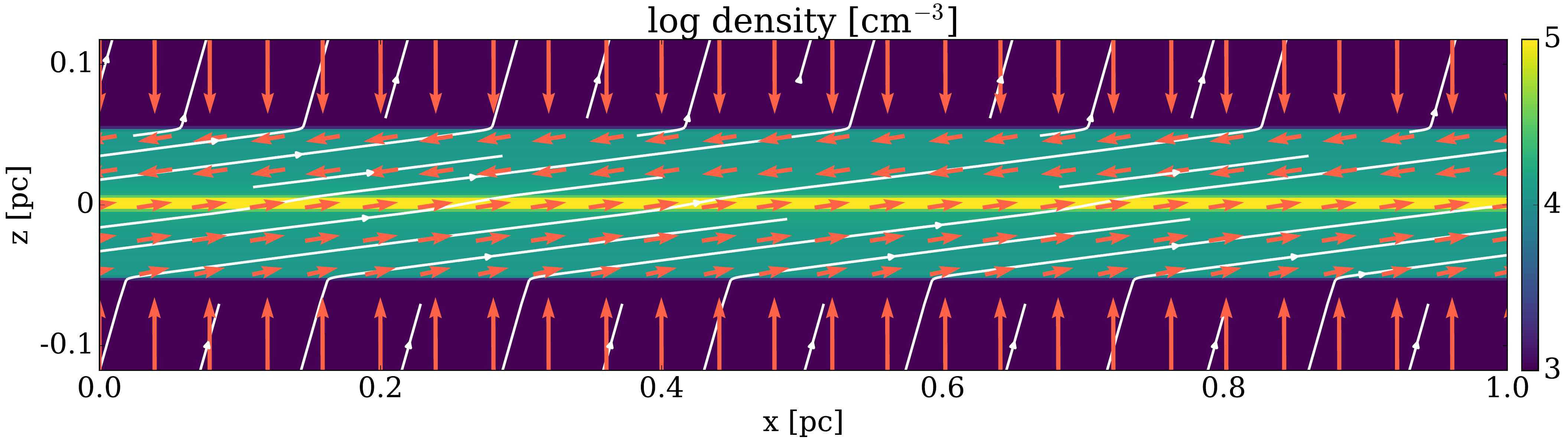}
\caption{Gas density ({\it colormap}), velocity ({\it red arrows}), and magnetic field ({\it white lines}) across a simulated oblique MHD shock system. In the post-shock region, material moves roughly parallel to the magnetic field lines, forming the secondary convergent flows that collide at the mid plane, forming a dense, sheet-like sub-layer.}
\label{sublayer}
\end{center}
\end{figure}

A plane-parallel oblique MHD shock from a numerical simulation with convergent inflow is illustrated in Figure~\ref{sublayer}. Within the post-shock layer, the magnetic field component parallel to the shock front is amplified by the shock, leading to an abrupt change of the magnetic field direction across the shock front. 
The sharp bending of magnetic field line provides a magnetic tension force that changes the direction of gas velocity, resulting in a non-zero velocity component parallel to the shock front. 
This can also be understood as follows: the post-shock medium is dominated by magnetic pressure in strong MHD shocks,\footnote{Under this assumption, the post-shock magnetic pressure is roughly equal to the inflow ram pressure: ${{B'}}^2/(8\pi) \sim \rho_0 {v_0}^2$. See \hyperlink{CO15}{CO15} for more detailed discussion.} and the post-shock gas flows are therefore confined by the magnetic field direction, which is oblique to the shocked layer because of the non-zero magnetic field component perpendicular to the shock front. 
These oblique post-shock gas flows will then collide at the central plane of the post-shock layer, further enhancing the density there. 

This {\it secondary convergent flow} is an important feature of oblique MHD shocks created by convergent flows, and its existence can be directly proven from MHD fluid dynamics and shock jump conditions. 
Since this feature has not been widely appreciated in the literature of cloud formation through converging magnetized flows, we will give a detailed derivation below. 

We will consider an oblique MHD shock with inflow velocity $\mathbf{v}_0 = v_0 \hat{\mathbf{z}}$, pre-shock magnetic field $\mathbf{B}_0 = B_{x0} \hat{\mathbf{x}} + B_{z0} \hat{\mathbf{z}}$, and uniform pre-shock gas density $\rho_0$.
The governing equations for MHD fluid are:
\begin{subequations}
\begin{align}
\frac{\partial\rho}{\partial t} &+ \mathbf{\nabla}\cdot\left(\rho\mathbf{v}\right) = 0,\label{goveqnmass}\\
%\rho\left[\frac{\partial v}{\partial t}  + \left(\mathbf{v}\cdot\mathbf{\nabla}\right) \mathbf{v}\right] - \frac{1}{4\pi}\left(\mathbf{\nabla}\times\mathbf{B}\right)\times\mathbf{B} & = 0,\label{goveqnmom}\\
\frac{\partial\rho\mathbf{v}}{\partial t} &+ \mathbf{\nabla}\cdot\left(\rho\mathbf{v}\mathbf{v} - \frac{\mathbf{B}\mathbf{B}}{4\pi}\right) + \mathbf{\nabla}\left(\rho {c_s}^2 + \frac{B^2}{8\pi}\right) = 0,\label{goveqnmom}\\
\frac{\partial\mathbf{B}}{\partial t} &+ \mathbf{\nabla}\times\left(\mathbf{B}\times\mathbf{v}\right) = 0,\label{goveqnB}
\end{align}
\end{subequations}
where we adopt an isothermal equation of state $P=\rho {c_s}^2$ with $c_s=0.2 \left(T/10\mathrm{K}\right)^{1/2}$~km/s. 
For the steady-state solution ($\partial_t = 0$) of a plane-parallel shock propagating along $z$ ($\partial_x = \partial_y = 0$), Equations~(\ref{goveqnmass}) and (\ref{goveqnmom}) give
\begin{subequations}
\begin{align}
\frac{d}{dz}&\left(\rho v_z\right) = 0, \\
\frac{d}{dz}&\left(\rho {v_x}{v_z} - \frac{B_x B_z}{4\pi}\right) = 0,\label{goveqnvx}\\
%\frac{d}{dz}&\left(\rho {v_y}{v_z} - \frac{B_y B_z}{4\pi}\right)  = 0,\\
\frac{d}{dz}&\left(\rho {v_z}^2 + \rho {c_s}^2 + \frac{{B_x}^2}{8\pi} \right) = 0.
\end{align}
\label{goveqndz}
\end{subequations}
and the Induction Equation (\ref{goveqnB}) becomes
\begin{equation}
\frac{d}{dz}\left(B_z v_x - B_x v_z\right) = 0.
\label{indjump}
\end{equation}
Also note that $d B_z /dz = 0$, since the magnetic field is divergence-free: $\mathbf{\nabla}\cdot \mathbf{B} = 0$. 
%Therefore ${B_z}' = B_{z0}$.
%Combining Equations~(\ref{goveqndz}) and (\ref{indjump}) gives the following jump conditions:
%\begin{subequations}
%\begin{align}
%\rho_0 v_0 &= \rho' {v_z}',\\
%-\frac{B_{x0} B_{z0}}{4\pi} &= \rho_0 v_0{v_x}'  -\frac{{B_x}' B_{z0}}{4\pi},\label{jumpvx}\\
%0 &= \rho_0 v_{z,0} {v_y}' -\frac{{B_y}' B_{z,0}}{4\pi},\\
%\rho_0 {v_0}^2 + \rho_0 {c_s}^2 + \frac{{B_{x0}}^2}{8\pi}& = \rho' {{v_z}'}^2 + \rho' {c_s}^2 + \frac{{{B_x}'}^2}{8\pi},\\
%B_{x0}v_0 &= B_{z0}{v_x}' - {B_x}' {v_z}'\label{jumpB}.
%\end{align}
%\label{jumpcdn}
%\end{subequations}

Following the convention introduced in \hyperlink{CO12}{CO12}, we define the post-shock density $\rho'$ as $\rho_0$ times the gas compression ratio $r$ ($\rho' \equiv \rho_0 r$, or equivalently ${v_z}' = v_0/r$ where ${v_z}'$ is the post-shock velocity component perpendicular to the shock front), the post-shock magnetic field component parallel to the shock front ${B_x}' \equiv B_{x0} r_B$ where $r_B$ is the magnetic compression ratio, and the shock parameters
\begin{equation}
\theta_0 \equiv \tan^{-1}\frac{B_{x0}}{B_{z0}},\ \ \ {\cal M} \equiv \frac{v_0}{c_s},\ \ \ {v_{\mathrm{A}_{x0}}}^2 \equiv \frac{{B_{x0}}^2}{4\pi\rho_0},
\end{equation}
which correspond to the angle between the pre-shock magnetic field and inflow, inflow Mach number, and upstream Alfv{\'e}n speed in the $x$-direction, respectively.
Equation~(\ref{indjump}) therefore gives
\begin{equation}
\frac{{v_x}'}{v_0} = \tan\theta_0\left(1+\frac{r_B}{r}\right),
\end{equation}
%which indicates the post-shock gas velocity is oblique to the initial inflow:
or
\begin{equation}
\tan\theta_v \equiv \frac{{v_x}'}{{v_z}'} = \frac{{v_x}'/v_0}{{v_z}'/v_0} = \left(r+r_B\right)\tan\theta_0.
\label{thetav}
\end{equation}
Since the compression ratios $r$ and $r_B$ are both positive numbers larger than $1$, $\theta_v$ is always nonzero in oblique shocks ($\theta_0 \neq 0$). Therefore, Equation~(\ref{thetav}) shows that a secondary flow is present in the post-shock region along the direction $\theta_v$ with respect to the original inflow direction. If the original shock is created by convergent flows, the post-shock layer would be bound by two oblique MHD shocks, with the flows immediately behind the two shocks moving in opposite directions, which inevitably leads to collision at the mid plane (see Fig.~\ref{sublayer}).  

We can similarly define
\begin{equation}
\tan\theta_B \equiv \frac{{B_x}'}{{B_z}'} = \frac{B_{x0} r_B}{B_{z0}} = r_B\tan\theta_0,
\label{thetaB}
\end{equation}
which describes the direction of the post-shock magnetic field. Since only the component perpendicular to the inflow is amplified during the shock, it is not surprising that $\theta_B > \theta_0$. 
Note that, comparing Equations~(\ref{thetav}) with (\ref{thetaB}) tells us that $\theta_v \geq \theta_B$ in oblique MHD shocks, which means gas does not flow perfectly along the magnetic field lines in the post-shock region. In fact, from Equations~(A10) and (A17) of \hyperlink{CO12}{CO12} (or Equations~(4) and (5) of \hyperlink{CO14}{CO14}), $r_B \sim r$ in strong MHD shock systems with $v_0/v_{\mathrm{A}x0} \gg 1$, which is indeed the environment mostly considered in star-forming process in MCs. 

\subsection{Central Sub-layer in the Post-shock Region}
\label{sec::sublayer}

\begin{figure}
\begin{center}
\includegraphics[width=\columnwidth]{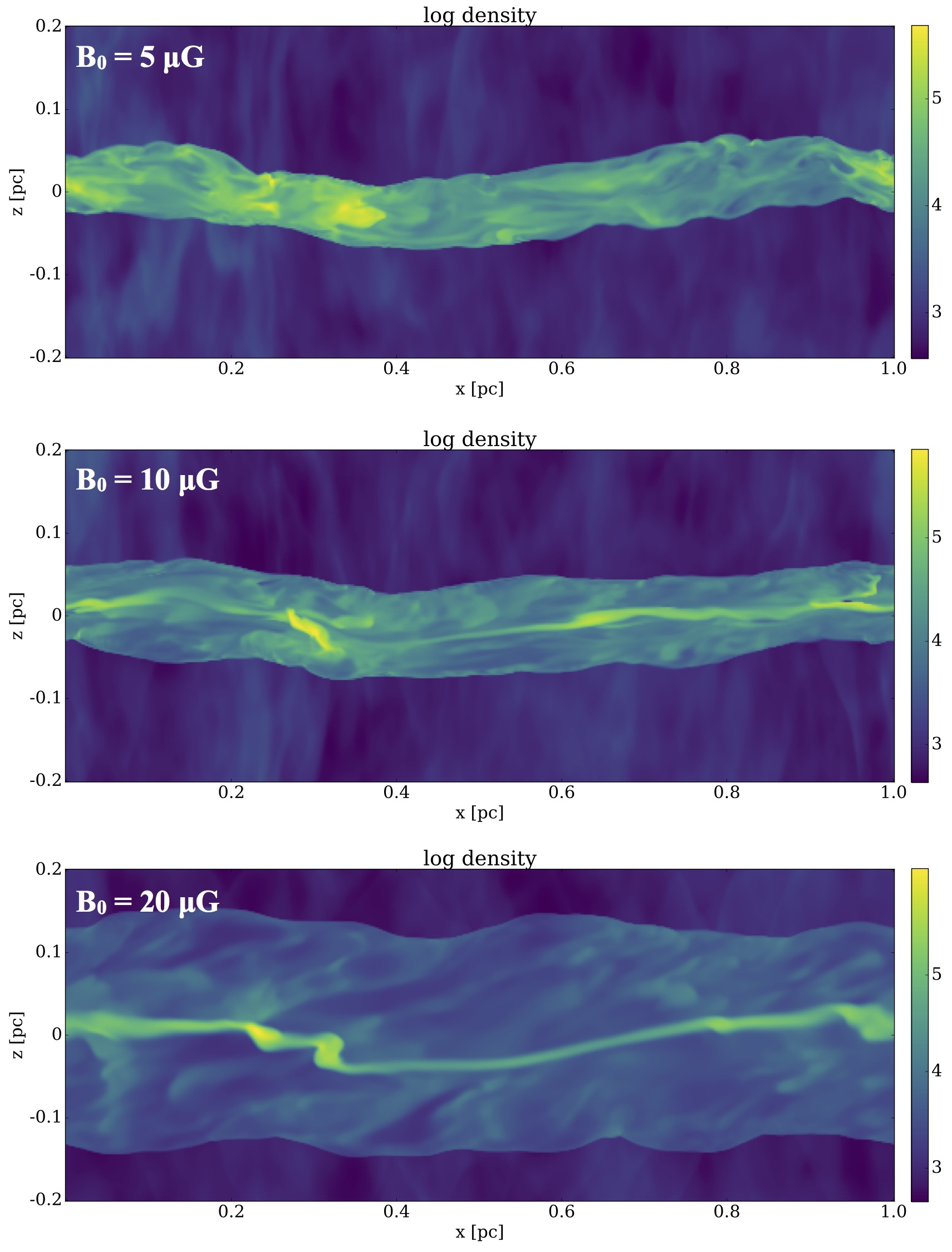}
\caption{Density slices of convergent-flow simulations discussed in CO15, with same initial conditions but different pre-shock magnetic field $B_0$, from models M10B5 ({\it top}), M10B10 ({\it middle}), and M10B20 ({\it bottom}). The dense sub-layer at the central plane of the post-shock region is more prominent when the pre-shock magnetic field is stronger ($B_0 = 10,\ 20~\mu$G). Note that since the post-shock region is perturbed by the turbulence in the inflow velocity, the dense sub-layer can only be clearly seen in density slices, not in integrated column density maps that are used in CO14. }
\label{sublayer_sim}
\end{center}
\end{figure}

In the simplest case of perpendicular shocks, the post-shock velocity ${v_z}'$ is likely very small because of a large jump in density. However, this does not mean that the post-shock flow is stagnated in oblique shocks, because with $r \gg 1$ and $r_B \gg 1$, Equation~(\ref{thetav}) gives ${v_x}' \gg {v_z}'$. The existence of large ${v_x}'$ hence significantly increases the strength of the secondary convergent flow. In addition, the angle between the post-shock velocity and magnetic field is greatly reduced from $90^\circ$ in the perpendicular shock to $(\theta_v - \theta_B)$ in the oblique shock, making it much easier to compress material in the central plane of the post-shock region, since the secondary collision is nearly along the field lines. 

To quantitatively demonstrate the strength of the secondary convergent flow relative to the original shock flow, we combine Equations~(\ref{goveqnvx}) and (\ref{indjump}) to derive\footnote{Note that Equation~(\ref{vx'}) is equivalent to Equation~(A7) in \hyperlink{CO12}{CO12}.}
%\begin{equation}
\begin{align}
\frac{{v_x}'}{v_0} & = \left(\frac{{B_{x0}}^2}{4\pi\rho_0{v_0}^2}\right)\left(\frac{B_{z0}}{B_{x0}}\right)\left(\frac{{B_x}'}{B_{x0}} - 1\right) \notag\\
&=\left(\frac{v_{\mathrm{A}x0}}{v_0}\right)^2 \frac{1}{\tan\theta}\left(r_B - 1\right),
\label{vxps}
\end{align}
%\end{equation}
or
\begin{equation}
\frac{{v_x}'}{c_s} = {\cal M}\left(\frac{v_{\mathrm{A}x0}}{v_0}\right)^2 \frac{r_B-1}{\tan\theta_0}.
\label{vx'}
\end{equation}

Since ${v_x}'$ is the dominant term of the secondary convergent flow, which is roughly parallel to the post-shock magnetic field, whether the secondary convergent flow will shock in the post-shock region will depend on if ${v_x}'/c_s \gtrsim 1$ or not. For strong shocks, ${\cal M}\gg 1$, $r_B \sim r \sim v_0 / v_{\mathrm{A}x0} \gg 1$ (see \hyperlink{CO12}{CO12} and \hyperlink{CO14}{CO14}), Equation~(\ref{vx'}) therefore becomes
\begin{equation}
\frac{v'}{c_s} \sim \frac{v_{\mathrm{A}x0}}{c_s} \frac{1}{\tan\theta_0} = \frac{v_{\mathrm{A}z0}}{c_s},
\label{v'}
\end{equation}
which suggests that secondary convergent flow can efficiently compress post-shock gas as shock waves if upstream magnetic field component perpendicular to the shock front is relatively strong so that $v_{\mathrm{A}z0} \gtrsim c_s$.
This is illustrated in Figure~\ref{sublayer_sim}, where we considered three models discussed in \hyperlink{CO15}{CO15} with different upstream magnetic field strength. While all other simulation parameters are kept the same, we clearly see the central sub-layer becomes more prominent with increasing upstream magnetic field $B_0$ (and hence $v_{\mathrm{A}z0}$).\footnote{Note that Equation~(\ref{v'}) only provides an order-of-magnitude estimate; though $v_{\mathrm{A}z0} / c_s \approx 0.8$ in the simulation considered in Section~\ref{sec:simstri} (model M10B10 of \hyperlink{CO15}{CO15}; middle panel in Figure~\ref{sublayer_sim}), the post-shock velocity in this model still reached supersonic values and therefore created a noticeable sub-layer.}
Equation~(\ref{v'}) also provides an explanation for the higher post-shock $v_\mathrm{rms}$ in models with stronger upstream magnetic fields presented in Table~1 of \hyperlink{CO15}{CO15}.

It may appear counter-intuitive that a substantial pre-shock field component perpendicular to shock front ($B_{z0}$) is needed to generate a supersonic post-shock flow parallel to the shock front (${v_x}'$). Physically, this is because the flow is generated by the magnetic tension of the sharply kinked field lines at the shock front, which increases with both the field strength and degree of field bending. Without a substantial perpendicular component ($B_{z0}$), the field lines would be nearly parallel to the shock front both before and after the shock, producing a kink too weak to accelerate the post-shock flow to a supersonic speed.

% !!! need more work!
In terms of star formation,
a dense sub-layer in the post-shock region could in principle shorten the timescale necessary to condense material to form self-gravitating structures.
Because of the compression from the secondary convergent flow, clumps within the sub-layer are born with higher volume densities (compared to clumps formed in systems without such a sub-layer). These clumps thus have better chances to collapse gravitationally in a relatively shorter timescale.
In addition, as we shall discuss in Section~\ref{sec::vx}, the shape of this dense sub-layer and its relative orientation to the local magnetic field are controlling the location of any condensed structure developed within it.

The dense sub-layer is a crucial factor for the formation of not only massive structures like prestellar cores and star-forming filaments, but also more diffuse striations. 
In the next section, we will demonstrate using existing simulations that magnetic field-aligned striations appearing in integrated column density map are likely the direct results of corrugations of the stagnated, sheet-like sub-layer. 
% !!!!!

\section{Simulated Striations}
\label{sec:simstri}

\subsection{Numerical Simulations}

Striations are commonly seen in numerical simulations. Here we use the core-forming simulations discussed in \hyperlink{CO14}{CO14} and \hyperlink{CO15}{CO15} to investigate the general features of simulated striations. These fully 3D MHD simulations are conducted using {\it Athena} \citep{2008ApJS..178..137S}, and are composed of $1$~pc-wide boxes with plane-parallel converging flows colliding along $z$-direction. The background magnetic field is on the $x$-$z$ plane, at $20^\circ$ to the $z$-axis. The oblique MHD shock hence creates a compressed post-shock layer with enhanced gas density, threaded by magnetic field approximately parallel to the $x$-axis (see e.g.~Figure~3 in \hyperlink{CO15}{CO15} and Figure~\ref{sublayer} above), wherein filamentary structures and prestellar cores form. The initial density is uniform at $1000$~cm$^{-3}$, and a randomly perturbed velocity field with power spectrum ${v_k}^2 \propto k^{-4}$ is applied to the entire simulation box. We adopted model M10B10 in \hyperlink{CO15}{CO15}, with inflow Mach number ${\cal M} = 10$ ($ v_0 = 2$~km/s at $T = 10$K) and cloud magnetic field $B_0 = 10~\mu$G. 

\begin{figure}
\begin{center}
\includegraphics[width=\columnwidth]{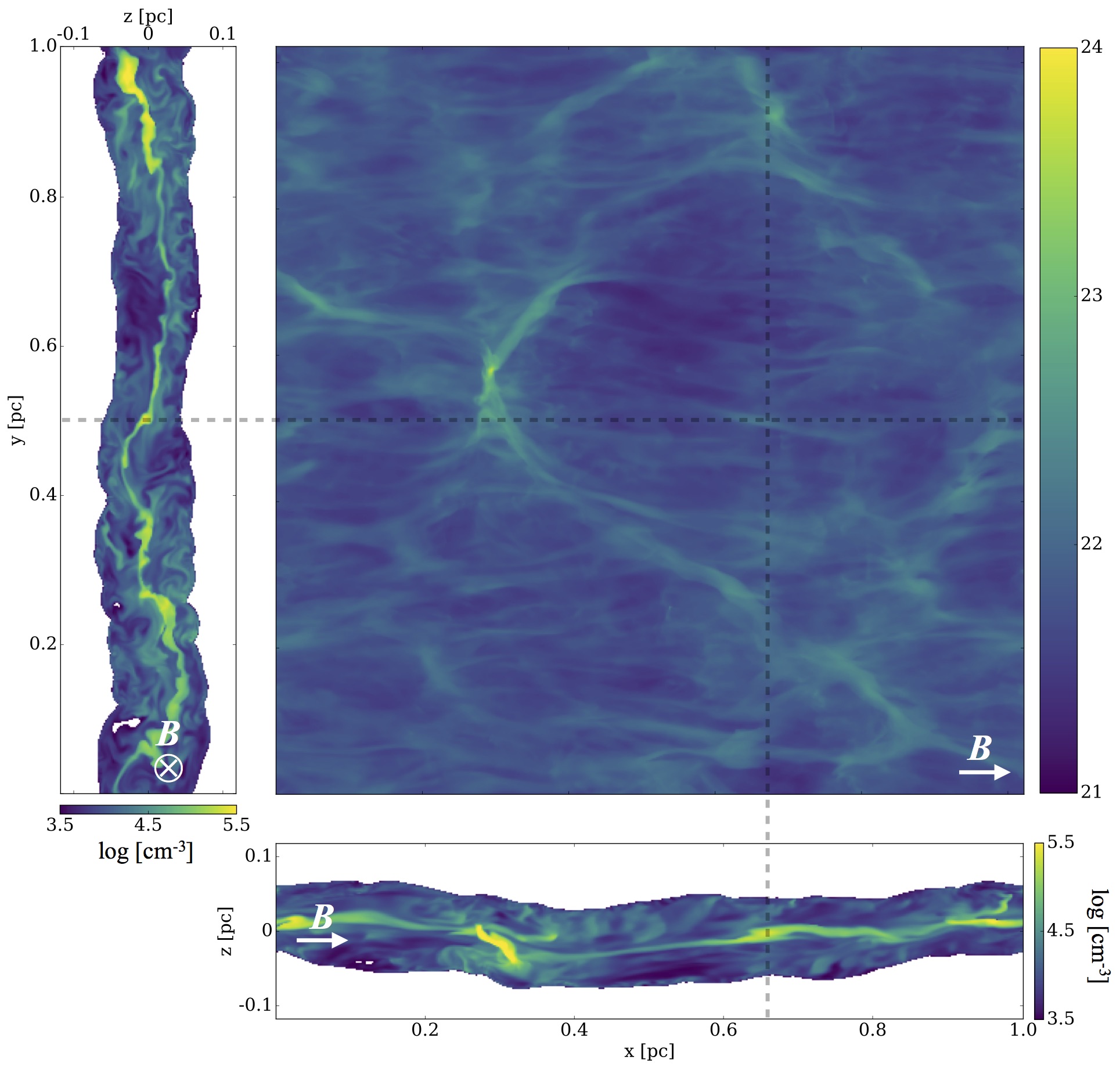}
\caption{Simulated column density map (in log scale, {\it top right}) of a $1\times 1$~pc$^2$ core-forming region generated by oblique MHD shocks, integrated along the plane-parallel convergent inflow ($z$-axis). Striations parallel to the magnetic field (whose approximate direction is indicated in each map) are clearly seen in the column density map. Two slices, cut through a striation at $x=0.64$~pc and $y=0.50$~pc ({\it dashed lines}), illustrate the density structures within the shocked layer approximately parallel ($x$-axis, {\it bottom}) and perpendicular ($y$-axis, {\it top left}) to the post-shock magnetic field. Along either viewing direction, there is a corrugated, thin, dense sub-layer at the mid plane of the shock. }
\label{M10map}
\end{center}
\end{figure}

Figure~\ref{M10map} shows the gas structures formed under this setup, at the time that the most evolved core starts to collapse ($t=0.51$~Myr in this particular simulation).\footnote{See \hyperlink{CO14}{CO14} and \hyperlink{CO15}{CO15} for justification of the choice of time frame.} The column density map (top right) integrated along $z$-direction shows the face-on view of the post-shock layer, while the two slices of gas density map (indicated by gray dashed lines) show the internal structures (through a striation; see Figure~\ref{sam2} for zoom-in maps) of the layer in $x$- and $y$- directions.\footnote{Note that the pre-shock region is removed from the maps and for all analysis on the core-forming simulation considered in this section and Section~\ref{sec::dis::core} to avoid possible contamination from the compressing inflows. For the simplified models discussed in Section~\ref{sec:turb} we kept the pre-shock regions to better show the full picture of the shock system.} 
The most evident feature in the post-shock region, as discussed in the previous section, is the dense sub-layer near the central plane, which is corrugated (compared to Figure~\ref{sublayer}) due to the velocity perturbation in the inflow gas. Note that this sub-layer contains more ripples along the $y$-direction while remaining relatively smooth along the $x$-direction (approximately parallel to the post-shock magnetic field). This motivates the further investigations to be found in Section~\ref{sec:turb}.
%By looking at the gas dynamics in the post-shock region, we realized this sub-layer is created by non-zero post-shock gas flows that are almost parallel to the magnetic field.  
%the secondary converging flows along the magnetic field lines in the post-shock region, and is a feature of oblique MHD shock; see Appendix~\ref{appdix} for an analytic derivation.

\subsection{General Properties}

\begin{figure}
\begin{center}
\includegraphics[width=\columnwidth]{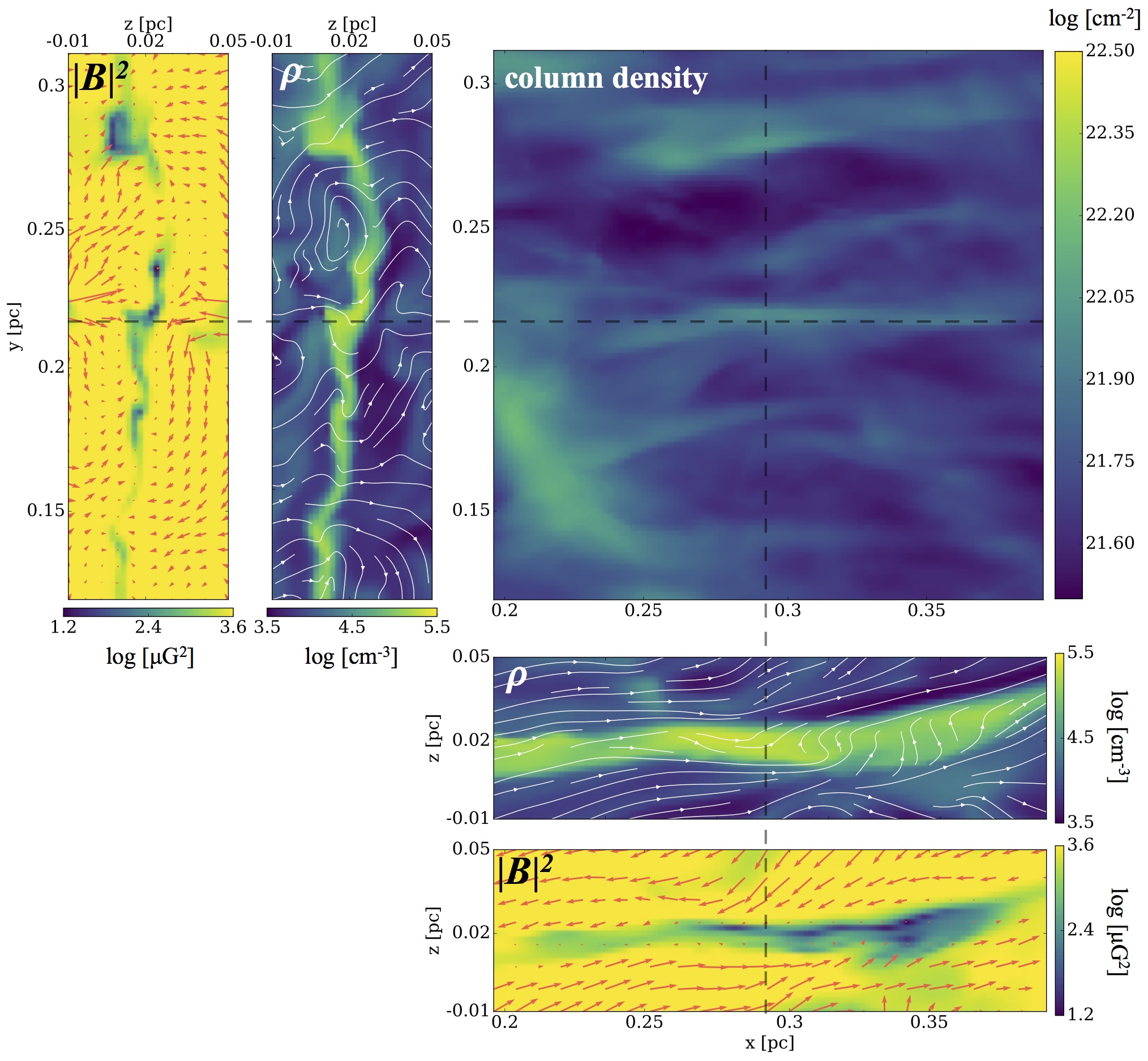}
\caption{A sample of simulated striations from the simulation shown in Figure~\ref{M10map}, in (a) column density integrated across the post-shock layer (along the large-scale inflow direction; in $\log($cm$^{-2})$, {\it top right}), (b) density $\rho$ (in $\log($cm$^{-3})$, {\it top middle} and {\it right middle}), and (c) magnetic pressure $|B|^2$ (in $\log(\mu$G$^2)$, {\it top left} and {\it bottom}). Density and magnetic pressure slices are cut through the gray dashed lines indicated in the column density map. White lines in the density maps indicate magnetic field directions, while red arrows in the magnetic pressure maps are gas velocities.}
\label{sam1}
\end{center}
\end{figure}

\begin{figure}
\begin{center}
\includegraphics[width=\columnwidth]{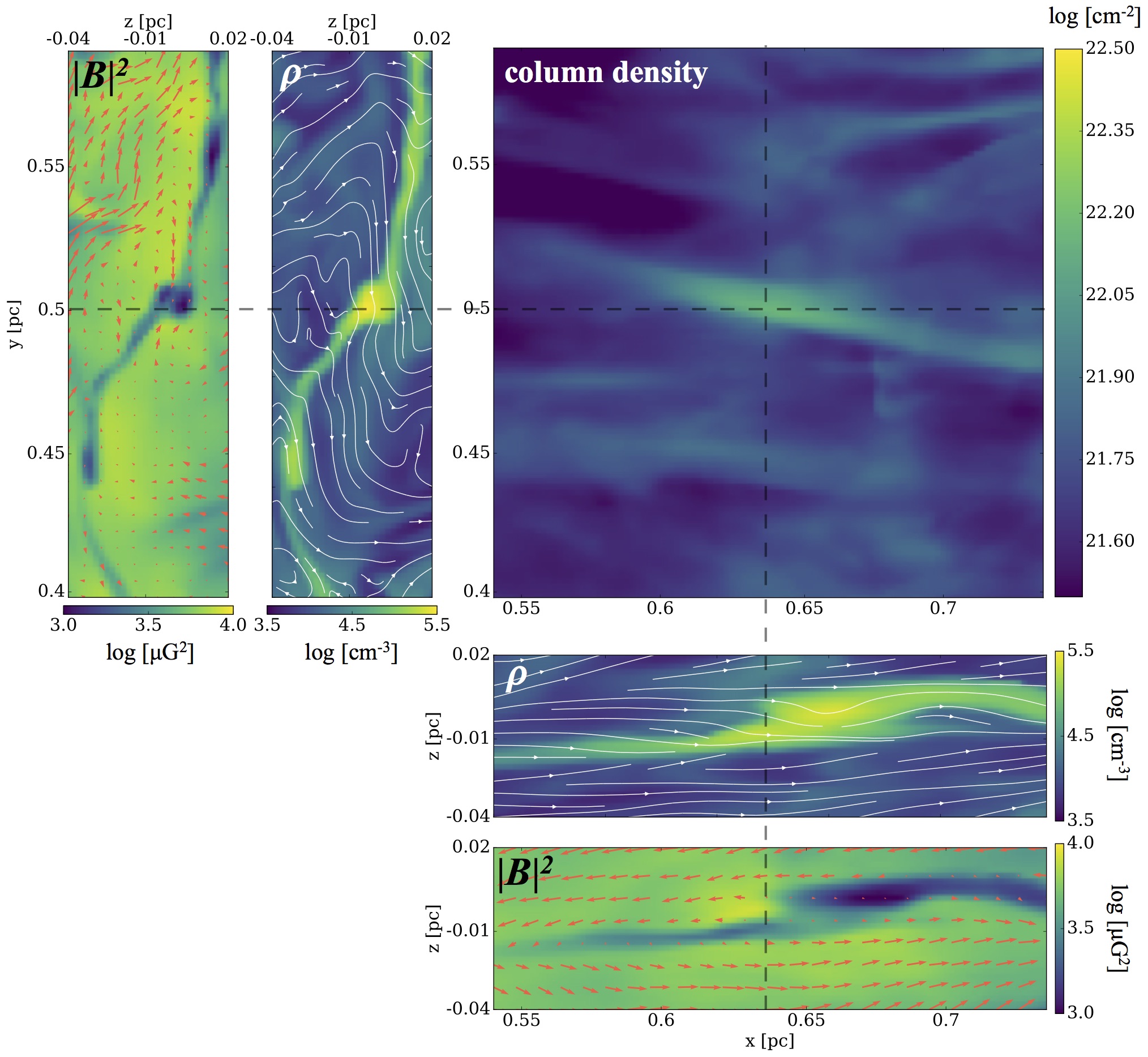}
\caption{Same as Figure~\ref{sam1}; another sample.}
\label{sam2}
\end{center}
\end{figure}

It is straightforward to show that filamentary structures and prestellar cores that form through collecting material along the magnetic field lines (see \hyperlink{CO14}{CO14} and \hyperlink{CO15}{CO15}) all lie within the central sub-layer, as well as the striations.
%Striations, though also locate in the sub-layer, align roughly parallel to the magnetic field and thus must have a different formation mechanism.
Figures~\ref{sam1} and \ref{sam2} illustrate two examples of the striations found in this simulation, showing both gas structure and magnetic pressure. 
One interesting feature is that the magnetic pressure tends to be lower within striations, despite the higher gas density. This indicates that these striations are not self-gravitating, which would result in enhanced magnetic field strength with higher density. In fact, the entire sub-layer is low in magnetization compared to the whole post-shock region, in order for the denser (and thus higher thermal pressure) sheet to be in pressure balance with its surroundings. This again shows that the dense sub-layer did not form from self-gravity but is a dynamic effect.

\begin{figure*}
\begin{center}
\includegraphics[width=\textwidth]{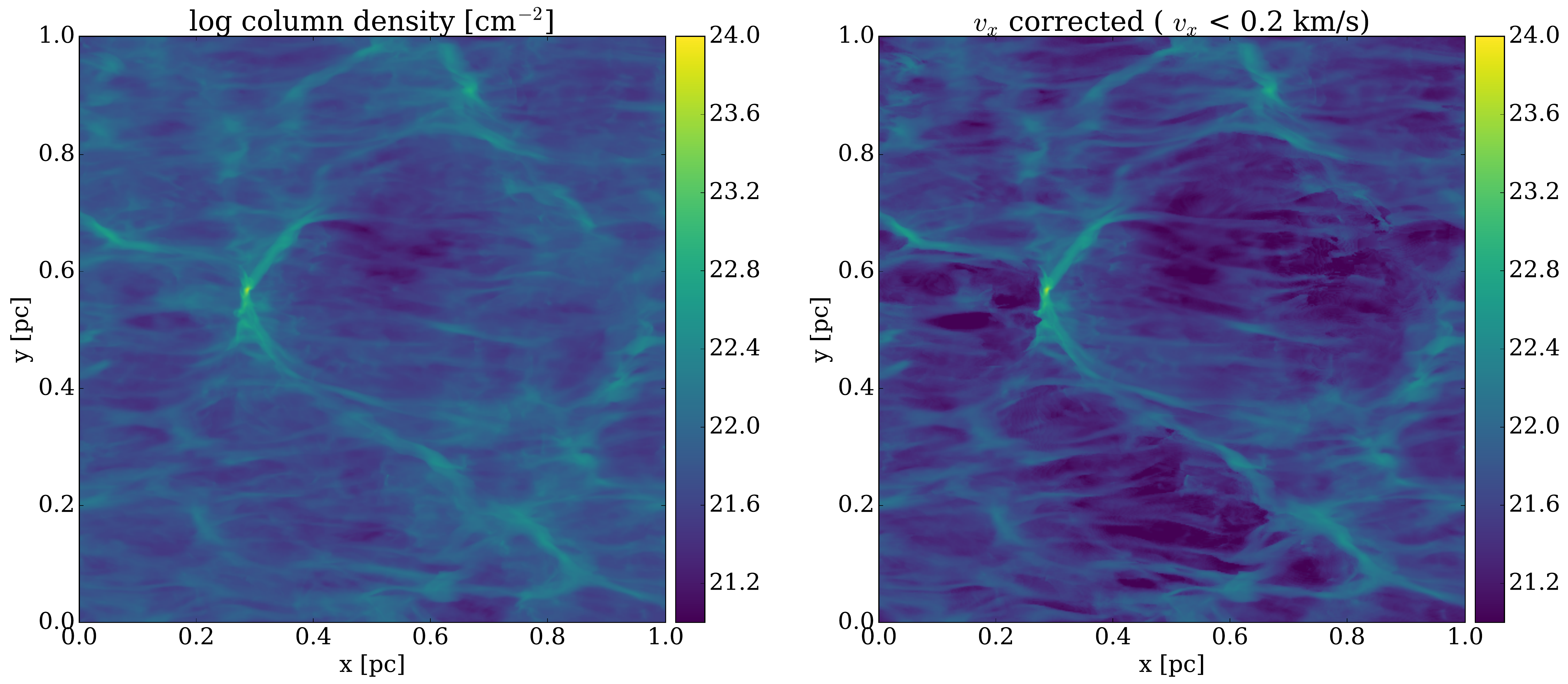}
\caption{The original column density map ({\it left}) comparing to the velocity-corrected map ({\it right}), which only includes cells with $v_x < 0.2$~km/s of the original simulation. Striations and filamentary structures are more prominent in the low-velocity map, which provides strong evidence that striations are within the stagnated sub-layer.}
\label{vxcorr}
\end{center}
\end{figure*}

In addition, we note that striations (and the dense sub-layer) have very low gas velocity, especially along the magnetic field. Since the dense sub-layer is created by collision of secondary convergent flows approximately parallel to the magnetic field, it is not surprising that $B$-direction gas velocity is low within the sub-layer. 
The low $B$-direction velocity in striations is a strong evidence that striations are stagnated features; they are not collecting material along the magnetic field gravitationally like star-forming filaments, nor are they leftover traces of gas feeding (moving towards) filaments. Indeed, if we consider only cells with low $v_x$ (since the post-shock magnetic field is approximately along $x$-direction under this setup; see e.g.~Figure~3 in \hyperlink{CO15}{CO15}) in the simulation and integrate along $z$ to make the same column density map as shown in Figure~\ref{M10map}, striations become more evident than the original map (Figure~\ref{vxcorr}). 
%We therefore conclude that striations are within the dense sub-layer and are stagnated gas structure that may be related to the corrugation of the thin layer.

More importantly, though density maps cut through the striations show that striations are associated with the dense sub-layer, we see that striations do not have significantly higher volume densities than their surroundings within the sub-layer. This is an interesting fact, because striations are, by definition, brighter regions in column density map.
On the other hand, striations seem to be associated with corrugated regions of the sub-layer that are apparent in density maps cut along the $y$ axis (see Figures~\ref{M10map}$-$\ref{sam2}). The longer path length of the corrugated layer explains the higher column density within striations. The questions now are: what is the physical process responsible for the corrugation of the sub-layer (and the formation of striations)? Why is the sub-layer more corrugated along the $y$-direction yet relatively smooth along $x$ (roughly parallel to the post-shock magnetic field)?

\section{Simplified Turbulent Models}
\label{sec:turb}

To understand why some parts of the sub-layer evolved to become striations and others did not, we ran similar simulations as discussed in \hyperlink{CO15}{CO15} with simplified turbulent models to investigate how velocity perturbation in the large-scale inflow affects the shape of the sub-layer and hence the formation of striations. In order to create predictable corrugations that can be easily interpreted physically, we consider the following form of perturbed inflow:
\begin{align}
v_z(x,y) = &-v_0\notag\\
& + \delta v_{z,x} \sin\left(2\pi\frac{n_x}{\ell_x}\right)\notag\\
& + \delta v_{z,y} \sin\left(2\pi\frac{n_y}{\ell_y}\right),\ \ \ z > 0\notag\\
v_z(x,y) &= v_0,\ \ \ z<0
\label{inflow}
\end{align}
where $n_x$, $n_y$ are grid numbers, and we set $\ell_x = \ell_y = N_\mathrm{box}/4 = 128$ for our $512^3$ simulations so there will be four complete cycles of sinusoidal ripples in $v_z$ along $x$- and/or $y$-direction within the simulation box. We select the perturbation amplitude in the original turbulence simulation, $\delta v = 0.14$~km/s, to be the value for $\delta v_{z,x}$ and $\delta v_{z,y}$, when applicable, so the simplified models are directly comparable to the original simulations.

\subsection{Variations Perpendicular to B and Magnetically Aligned Striations}
\label{sec::vy}

\begin{figure}
\begin{center}
\includegraphics[width=\columnwidth]{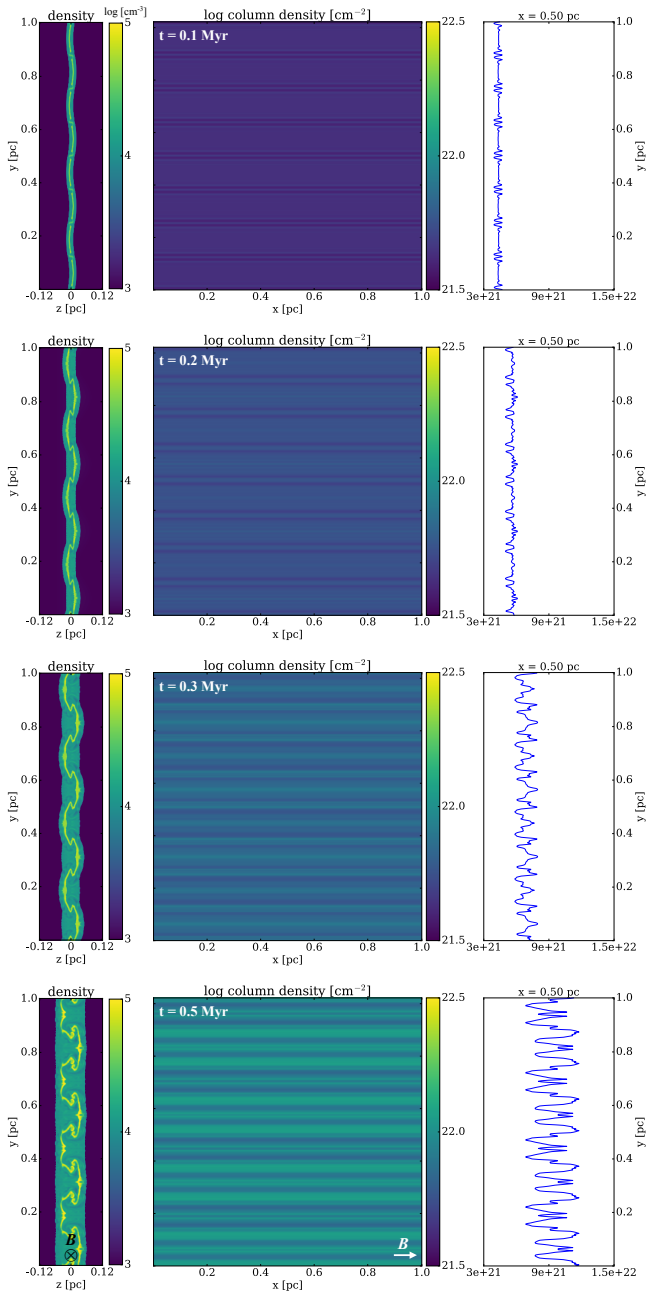}
\caption{Simulation with velocity perturbation along $y$-direction (perpendicular to the magnetic field, marked on the bottom panels), presented in gas density cut through $x=0.5$~pc ({\it left panel}), integrated column density across the post-shock region ({\it middle panel}), and line plot of the column density ({\it right panel}).}
\label{sinvy}
\end{center}
\end{figure}

We first illustrate the basic mechanism for the formation of striations using inflow velocity variations along the $y$-direction ($\delta v_{z,y} = \delta v$, $\delta v_{z,x} = 0$), which is perpendicular to the magnetic field both upstream and downstream. The evolution of gas structure in this setup is shown in Figure~\ref{sinvy}; thin, striation-like stripes parallel to $x$ (the dominant magnetic field direction in the post-shock region) can be found everywhere in the column density map of this simulation (middle panel of Figure~\ref{sinvy}). 

Looking at the density structure of a slice cut through $x=0.5$~pc (left panel of Figure~\ref{sinvy}), we see a sinusoidal post-shock layer at the very beginning of the simulation (frame 1; $t=0.1$~Myr) as well as a dense sheet at the center. 
% !!! need to clarify
This thin sub-layer, however, becomes more and more corrugated as the simulation progresses (frame 2$-$4; $t=0.2-0.5$~Myr).
%\footnote{The wrinkled sheet seems to induce the smoothing of the shape of the post-shock layer; see Figure~\ref{sinvx} for a comparison.}
% !!!
Viewed along the inflow direction, these crumpled regions appear as striation-like horizontal structures parallel to the magnetic field in the column density map (middle panel).

To better understand the column density structure of these dense stripes, the line plot of the column density at $x=0.5$~pc is also shown in Figure~\ref{sinvy} (right panel). These plots suggest that small perturbations at early times gradually evolve to prominent filamentary features with larger column-density contrasts, which resemble the striations we see in both observations and original turbulence simulations.
We note that these enhancements in column density correspond to certain locations of the sine-wave velocity variation; specifically, they occur where the rippled sub-layer has the steepest slope, either positive or negative, with respect to the vertical $y$ axis (e.g.~$y=0,\ 64,\ 128$ of the $n_y=0-128$ sine wave cycle). Though not necessarily regions of highest volume density, corrugations at these locations are further enhanced during the simulation (see the left panel of Figure~\ref{sinvy}).

All of these features can be explained by the non-linear thin shell instability (NTSI) in hydrodynamics, first discussed by \cite{1994ApJ...428..186V}. The NTSI takes place exclusively in corrugated, shock-compressed layers, where the ram pressure of the inflow is misaligned with the thermal pressure of the shocked gas (see Figure~1 of \citealt{2013MNRAS.431..710M}). In pure hydrodynamics, this imbalance results in a shearing flow along the corrugated sheet, accumulating materials at local extremities of the ripples and thus enhancing corrugation. This effect has been reported in many numerical simulations of collisions between molecular gas \citep{1996NewA....1..235B,1998ApJ...497..777K,2003NewA....8..295H,2013MNRAS.431..710M} or unmagnetized plasma \citep{2015PhRvE..92c1101D,2017MNRAS.465.4240D,2017ApJ...834L..21A}.

In the presence of a magnetic field, the growth of NTSI can be different, especially for perturbations that would lead to significant field bending, which would be resisted by the magnetic tension force.\footnote{Note that the magnetized NTSI was investigated by \cite{2007ApJ...665..445H}. However, their main focus was on the effect of magnetization on the NTSI growth rate, and their 2-D model cannot capture the folding/bending effect of the thin sheet around magnetic field threading the sheet.} Since the perturbations in the simplified turbulence model under consideration are spatially uniform along the $x$-direction (roughly the post-shock field direction) they do not produce significant bending of the magnetic field lines, and are thus not inhibited by magnetic tension. They grow in amplitude with time and produce shearing forces that roll up the dense sheet (see left panels of Figure~\ref{sinvy}) around the major direction of the magnetic field threading the sheet.  Since the shearing force is stronger where the inflow and the sheet are at a larger angle, such regions tend to be more strongly perturbed. 
%In a highly-magnetized environment as the one discussed here, the magnetic field in the post-shock region is strong enough to prohibit gas movement on the $y$-$z$ plane that would lead to field bending. The shearing flow and growth rate of the instability is therefore suppressed; instead, the shearing force rolls up the dense sheet (see left panel of Figure~\ref{sinvy}) around the magnetic field threading it so that materials do not move across magnetic field lines.
%Such motions do not lead to the bending of the field lines, and thus are not suppressed by magnetic tension. 
%Since the shearing force is stronger where the inflow and the sheet have a larger angle (more unbalanced), those regions in the dense sub-layer with steeper slopes tend to be more perturbed.

These wrinkled, folded regions thus have higher column densities when integrated across the sub-layer (perpendicular to the primary shock front, along the large-scale inflow), and become those field-aligned, striation-like stripes that we see on the column density map (middle panel of Figure~\ref{sinvy}). 
%Note that the magnetized NTSI was investigated by \cite{2007ApJ...665..445H}; though their main focus was on the effect of magnetization on the NTSI growth rate, similar folding/bending effect can be clearly seen in their simulations (e.g.~their Figure~8, left panel), which strengthens the case that the folding is a general feature of colliding flows. Although not commented on by the authors, 
Therefore, we believe the folding effect of the dense sub-layer around magnetic field within the primary post-shock region created by large-scale inflow is a key ingredient for striation formation in interstellar clouds.

\begin{figure}
\begin{center}
\includegraphics[width=\columnwidth]{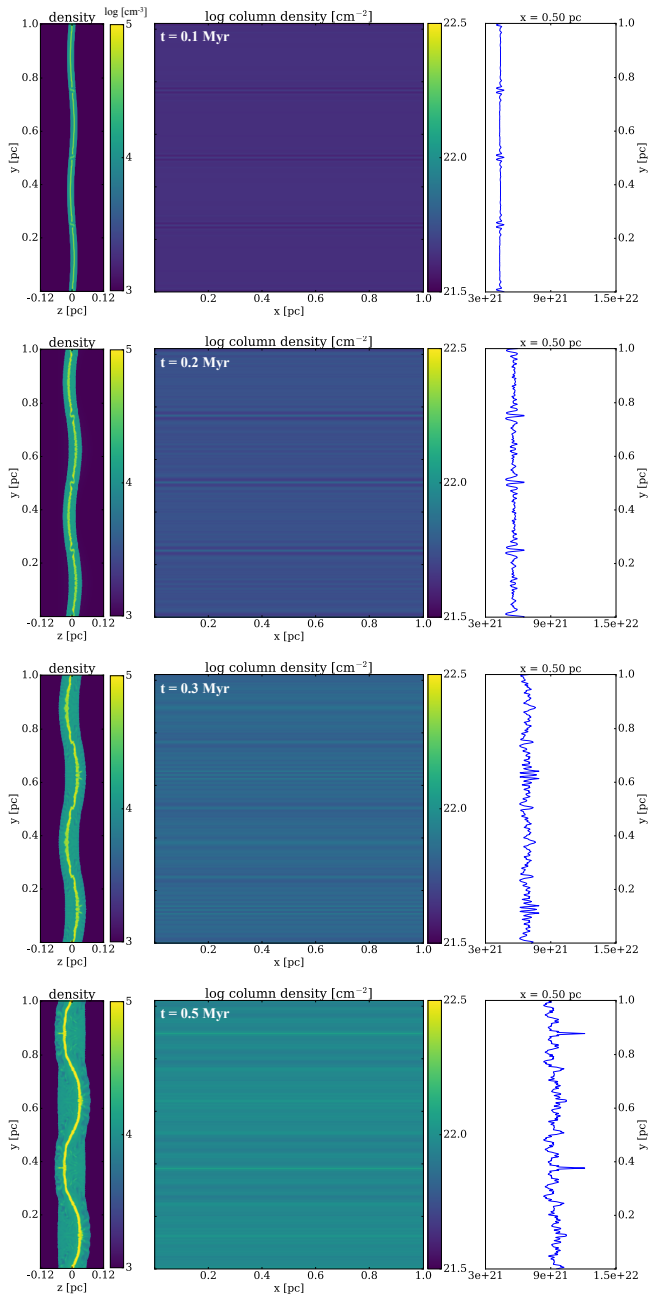}
\caption{Same as Figure~\ref{sinvy}, but showing the simulation with longer-wavelength inflow velocity variation, $\ell_y = N_\mathrm{box}/2$. The dense sub-layer is less corrugated from the shearing force, resulting in less-prominent striations.}
\label{sinLvy}
\end{center}
\end{figure}

To test the idea that the striations formed in this simulation are the direct results of NTSI, we repeated the simulation with same setup but a longer wavelength of the sinusoidal variation in inflow velocity ($\ell_y = N_\mathrm{box}/2 = 256$ in Equation~(\ref{inflow})), to see how the locations of the striations are affected. The results are shown in Figure~\ref{sinLvy}; we again see a rippled sub-layer with small perturbations at locations with steepest slopes (left and right panels of Figure~\ref{sinLvy}, frame 1 and 2). However, in this model the shearing force is too weak to significantly fold/roll up the magnetized sub-layer, and therefore the striations in this simulation are less bright compared to the previous one (middle panel of Figure~\ref{sinLvy}). Note that this setup gives a result more similar to the hydrodynamic NTSI in that local extremities grows to higher densities from shearing flows (see frame 4 of Figure~\ref{sinLvy}).

\begin{figure*}
\begin{center}
\includegraphics[width=\textwidth]{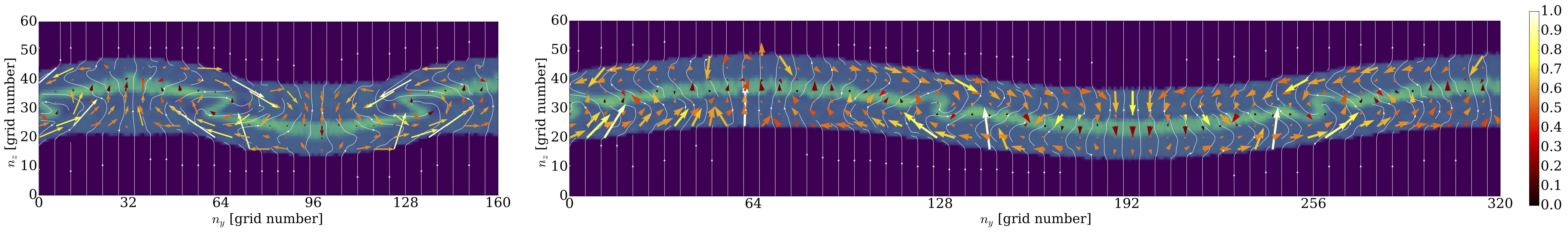}
\caption{Slices of gas density (in log scale, {\it colormap}), velocity (in km/s, {\it color-coded arrows}), and magnetic field ({\it white lines}) structures in simulations with different inflow velocity variation: shorter ({\it left}, also see Figure~\ref{sinvy}) and longer ({\it right}, also see Figure~\ref{sinLvy}) wavelengths in the direction perpendicular to the magnetic field, at $t = 0.3$~Myr. When the post-shock region is more curved in the beginning (shorter-wavelength perturbation), the shearing force from NTSI is more significant to induce corrugation of the dense sub-layer around the major direction of the post-shock magnetic field (pointing into the plane of the paper; not shown).}
\label{vyLvy}
\end{center}
\end{figure*}

Figure~\ref{vyLvy} demonstrates detailed gas dynamics of these two cases. In the shorter-wavelength model (left panel), the imbalance between ram pressure and thermal pressure is more significant due to the larger angle/steeper slope between the dense sub-layer and the inflow, and therefore leads to more robust folding of the sheet and rolling up of the magnetic field. On the other hand, the model with the longer perturbation wavelength (right panel) has a less-curved post-shock layer, and thus the shearing force is perturbing the gas/magnetic structure within the sub-layer to a lesser extent.

\subsection{Variations along B}
\label{sec::vx}

\begin{figure*}
\begin{center}
\includegraphics[width=\textwidth]{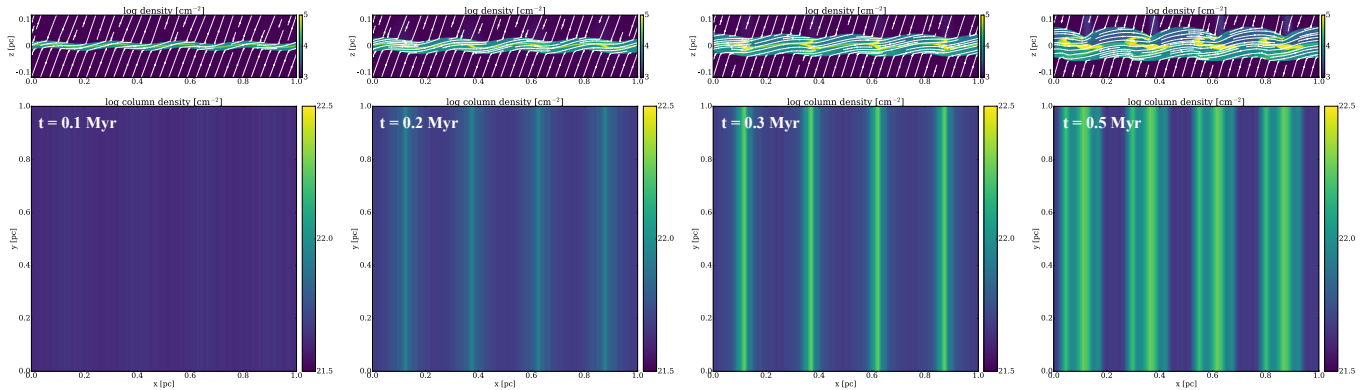}
\caption{Simulations with sinusoidal perturbation along $x$ (roughly parallel to the post-shock magnetic field). {\it Top panel:} gas density ({\it colormap}) and magnetic field structure ({\it white lines}) from a slice cut through $y=0.5$~pc. The stagnated sub-layer can be easily seen in the mid plane of the post-shock region. {\it Bottom panel:} column density map integrated along $z$ (primary inflow direction). Some ``preferred locations" of higher column densities have emerged since very early stages of the simulation (see Figure~\ref{vxconv}).}
\label{sinvx}
\end{center}
\end{figure*}

Besides diffuse, field-aligned striations, the dense sub-layer can also produce massive filaments tilted significantly away from the magnetic fields under certain conditions. We illustrate filament formation with plane-parallel inflows along the $z$-direction with velocity variation in the $x$-direction ($\delta v_{z,x} = \delta v$, $\delta v_{z,y} = 0$) so the perturbation induces variation roughly along the post-shock magnetic field.
The results are shown in Figure~\ref{sinvx}, with both the evolution of the post-shock region (slices of gas density in log scale at $y=0.5$~pc) and the corresponding column density maps integrated along the inflow (face-on view of the post-shock layer). 
% !!!! need to clarify
Similar to Figure~\ref{sinvy}, a sinusoidal post-shock layer appears in the first frame ($t = 0.1$~Myr), which reflects the spatial variation of the inflow velocity (four complete sine waves within the box; see Equation~(\ref{inflow})).
%The initial corrugation can be seen in the first frame ($t = 0.1$~Myr), that the post-shock region is shaped by the spatial variation of the inflow velocity. 
% !!!!
Also, the dense sub-layer compressed by the secondary converging flow has become prominent at the central plane of the post-shock region since the second frame ($t = 0.2$~Myr). 

In addition to the existence of the sub-layer, some ``preferred locations" emerge, forming repeated filamentary structures vertically along $y$ at certain $x$ locations with fixed separation (see the bottom panel of frame 2). At these locations, density increases over time (frame 3; $t = 0.3$~Myr), presumably by collecting material via the secondary flow along the magnetic field. Though the tension of the post-shock magnetic field is restricting further bending of the dense sub-layer in the post-shock region (see frames 3 and 4 of the top panel), 
enough mass can accumulate at each of these ``preferred locations" for the gas self-gravity to draw in more material (frame 4; $t=0.5$~Myr). This, similar to the anisotropic formation model described in \hyperlink{CO14}{CO14} and \hyperlink{CO15}{CO15}, is a way to produce overdense filaments slanting away from the magnetic field (rather than the field-aligned striations).\footnote{Note that, as discussed in \hyperlink{CO15}{CO15}, filaments formed via anisotropic condensation are not necessarily perpendicular to the magnetic field (i.e.~concentrating symmetrically along the field line on both sides of the filament), because the intrinsic gas velocity is not likely uniform around the filament. }
% !!!!! Can you be sure that the filaments are indeed self-gravitating at late times? 

\begin{figure}
\begin{center}
\includegraphics[width=\columnwidth]{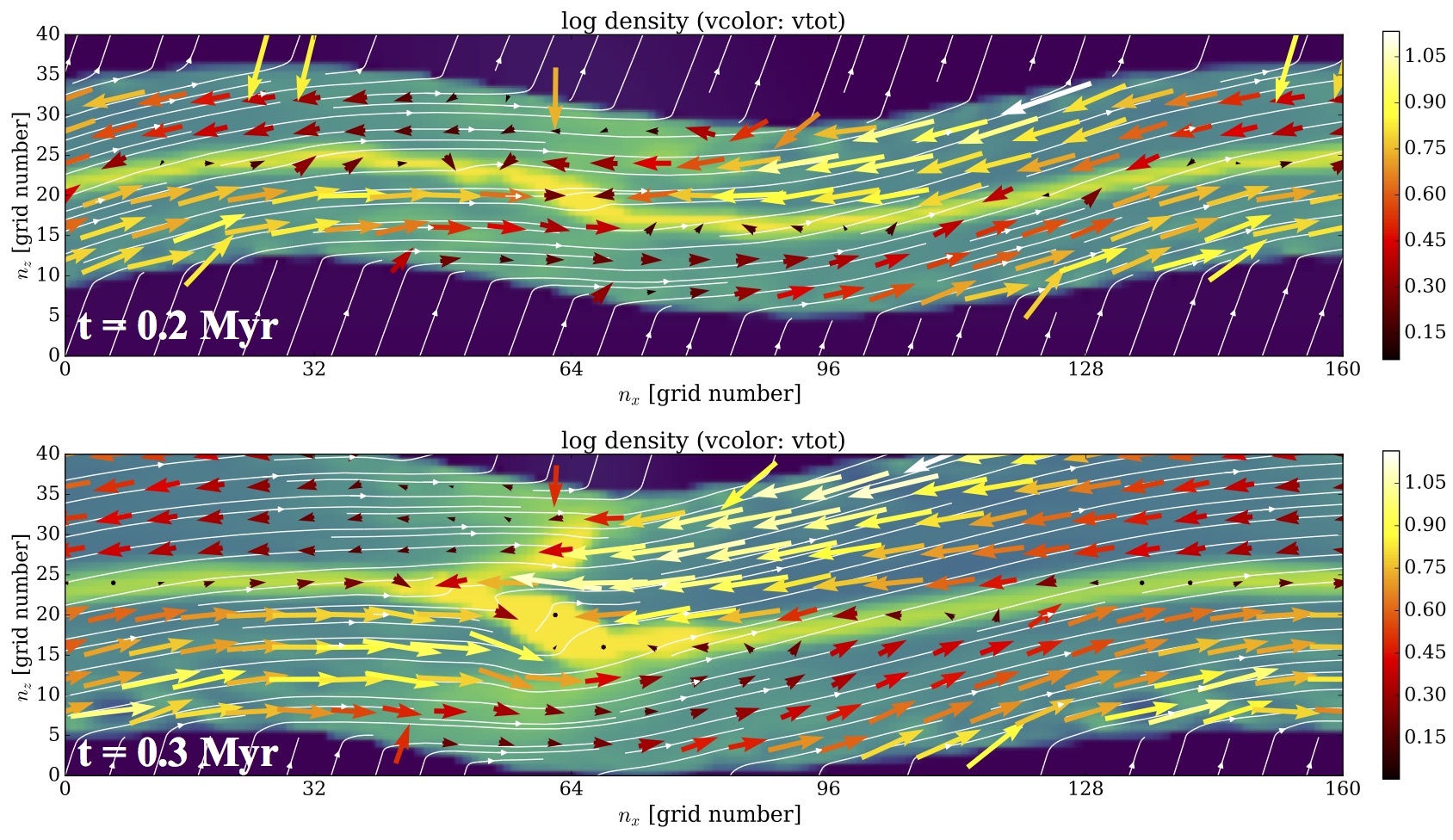}
\caption{The secondary convergent flow ({\it colored vectors}) in corrugated post-shock region ({\it colormap} of density in the background, with same color scale as in the top panel of Figure~\ref{sinvx}). Ripples with negative slopes ($n_x \approx 32 - 96$ in the sine wave cycle of $128$) are at larger angles with respect to the local magnetic field ({\it white lines}), and therefore have stronger convergence from the gas flows approximately along the magnetic field. These ``preferred locations" may be seeds of filaments or cores in the turbulent simulations. }
\label{vxconv}
\end{center}
\end{figure}

The slanted overdense filaments are formed where the secondary flows in the post-shock region are converging more strongly. The compressing strength of the secondary converging flows is determined by the angle between the sub-layer and the local magnetic field. Figure~\ref{vxconv} illustrates the detailed gas velocity and magnetic field within a zoomed-in region of the post-shock slab, at a time before self-gravity became significant. As we already saw in Figure~\ref{sinvx}, the shape of the initial ripples (for both the post-shock slab and the dense sub-layer) generally follows the sinusoidal form of the initial perturbation, which gives a full cycle between grid number in the $x$ direction $n_x = 0 - 128$ (see Equation~(\ref{inflow})).
% !!! need to clarify
However, since the non-zero magnetic field perpendicular to the shock front ($B_z$) is not altered during the shock, the post-shock magnetic field always has a positive slope (upper-right to bottom-left) on the $x$-$z$ plane (see Figure~\ref{vxconv}).
As a result, the post-shock magnetic field is not as crumpled as the gas, and
% !!!
sections of the sub-layer tilted with negative slopes (upper-left to bottom-right; $n_x \approx 32 - 96$) are at larger angles with respect to the magnetic field than other sections, especially those with positive slopes ($n_x \approx 96 - 160$). This, combined with the fact that post-shock gas flows are mostly along the magnetic field lines, determines where the secondary convergent flow results in the strongest compression, forming those overdense filaments aligned vertically (perpendicular to the post-shock magnetic field) at certain $x$ locations in Figure~\ref{sinvx} that can condense further gravitationally. 

\subsection{Combined variations}
\label{sec::vxy}

\begin{figure}
\begin{center}
\includegraphics[width=\columnwidth]{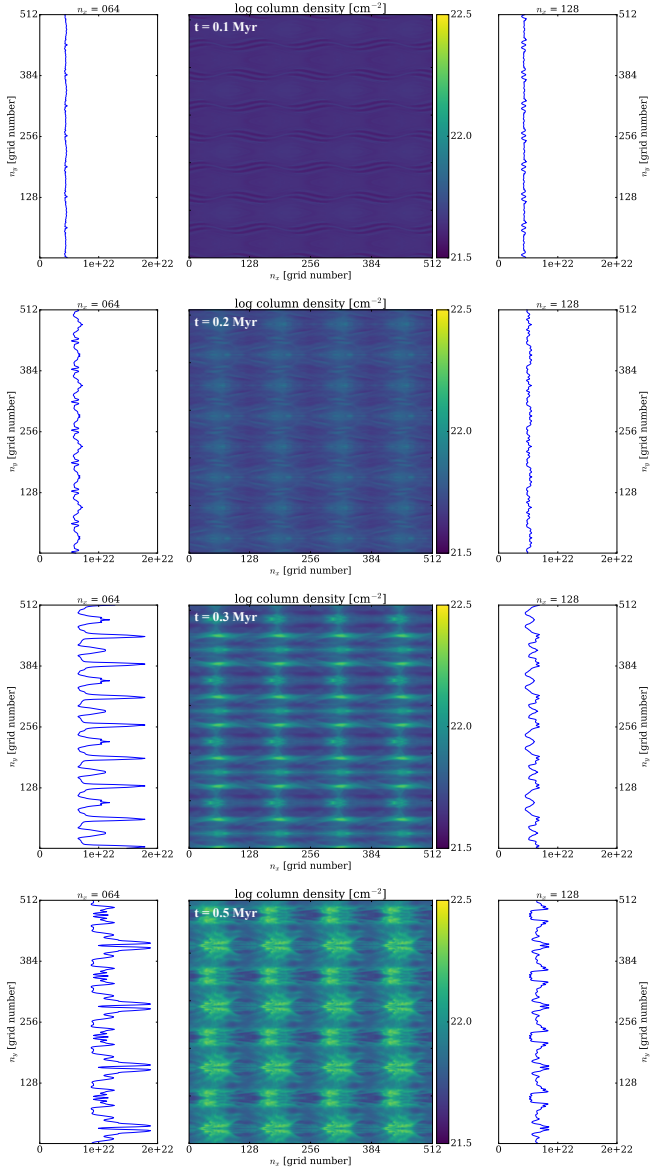}
\caption{Simulation with inflow variations at both parallel and perpendicular to the post-shock magnetic field; $\delta v_{z,x} = 2~\delta v$, $\delta v_{z,y} = 0.5~\delta v$ in Equation~(\ref{inflow}). Striation-like structures are clearly seen in the column density maps ({\it middle panel}), while the evolution of column density along $y$ can be very different at $x=64$ (the ``preferred location" where secondary convergent flow is the most efficient in the dense sub-layer; {\it left panel}) and $128$ (the no-collision point where less or no gas flows collide; {\it right panel}) because of the shape of the corrugated post-shock layer (see Figure~\ref{vxconv}).}
\label{sinvwx}
\end{center}
\end{figure}

After investigating inflow variations parallel and perpendicular to the magnetic field separately, we ran an additional simulation to see how gas structures form under the combination of these two effects. To emphasize the creation of overdense structures through flows along the magnetic field, we considered $\delta v_{z,x} = 2~\delta v$, $\delta v_{z,y} = 0.5~\delta v$ to slightly reduce the significance of NTSI. 

The results are shown in Figure~\ref{sinvwx}; striations are clearly seen in the column density map (middle panel), especially in frame 3 ($t=0.3$~Myr) before the gas clumps become too massive. The ``preferred locations" as discussed in Section~\ref{sec::vx} can also be easily identified from the column density map.\footnote{Note that in Figure~\ref{sinvwx} we use grid numbers instead of physical length scales for both $x$ and $y$ axes for the readers to easily locate the sine wave cycles.} To see how gas structures evolve under MHD NTSI, we compare the line plots of column density along $y$-direction at two different $n_x$ values:
% !!! need to clarify
one at the ``preferred location" ($n_x = 64$; left panel) where the secondary flow has the largest angle with respect to local magnetic field, and one at the no-collision point ($n_x = 128$; right panel) where the gas flow is almost parallel to field lines (see Figure~\ref{vxconv}). 
% !!!

Though the initial perturbation level is about the same at these two locations (frame 1 and 2), gas structures at the ``preferred locations" quickly grow into self-gravitating clumps, reaching relatively high column density (frame 3 and 4 in the left panel). On the other hand, at the no-collision point (right panel) there is no significant material flowing toward it, and thus the overall column density remains at an average level, keeping these striations just slightly distinguishable from the background.

\section{Discussions}
\label{sec::disc}

\subsection{Formation Criteria of Striations}

As mentioned in Section~\ref{sec:intro}, computational studies have suggested that prominent striations only appear in media with moderately strong magnetization (\citealt{2008ApJ...680..420H}; \hyperlink{CO14}{CO14}; \hyperlink{CO15}{CO15}). Indeed, this can be explained by the formation criterion of the dense sub-layer discussed in Section~\ref{sec::sublayer}; Equation~(\ref{v'}) gives a rough estimate of the minimum strength of the pre-shock magnetic field associated with the creation of a secondary convergent shock. In our scenario that the presence of the dense sub-layer is crucial to the formation of striations (because shearing forces would not be dynamically efficient without such a thin sub-layer), Equation~(\ref{v'}) suggests that striations only exist in strongly magnetized post-shock regions.

From previous simulations, it has been estimated that the condition for striations to exist is $\beta \lesssim 0.2$ for ${\cal M} \sim 5$ \citep[see e.g.][]{2008ApJ...680..420H}. Considering the definition of plasma beta $\beta \equiv 8\pi\rho {c_s}^2 / (B^2)$, its value in the post-shock region for dense sub-layer and striations to exist can be written as (following the same convention in Section~\ref{scf} and assuming the balance between the upstream ram pressure and the downstream magnetic pressure)
\begin{equation}
\beta' = \frac{8\pi \rho' {c_s}^2}{{B'}^2} \sim \frac{\rho' {c_s}^2}{\rho_0 {v_0}^2} = \frac{r}{{\cal M}^2}.
\end{equation}
Since $r \sim v_0 / v_{\mathrm{A}x0} = {\cal M}/{\cal M}_{\mathrm{A}x0}$ (see Section~\ref{sec::sublayer}, or see \hyperlink{CO14}{CO14}, \hyperlink{CO15}{CO15} for full derivation), the criterion for striations to emerge becomes
\begin{equation}
\beta' \sim \frac{1}{{\cal M}{\cal M}_{\mathrm{A}x0}} \lesssim 0.2,
\end{equation}
or equivalently,
\begin{align}
{\cal M}_{\mathrm{A}x0} &= {\cal M}_{\mathrm{A}z0} \tan\theta_0 \gtrsim \frac{1}{0.2~{\cal M}},\notag\\
{\cal M}_{\mathrm{A}z0} & \gtrsim \frac{1}{0.2~{\cal M}\tan\theta_0 }.
\end{align}
This roughly agrees with Equation~(\ref{v'}), which requires ${\cal M}_{\mathrm{A}z0} = v_{\mathrm{A}z0} / c_s \gtrsim 1$.

\subsection{Anisotropic Corrugations}

By conducting simulations with simplified velocity perturbation models in the inflow, we have clarified the role played by gas dynamics in shaping the dense sub-layer in the post-shock region, and distinguished between the effects of inflow velocity variation parallel and perpendicular to the post-shock magnetic field. 
In the case of a perturbation in inflow velocity varying along the post-shock magnetic field (i.e.~along the $x$-direction in our setup), the dense sub-layer and the post-shock magnetic field lines that guide the secondary converging flows to form the sub-layer in the first place respond differently to the perturbation. Because of their rigidity in the mostly sub-Alfv{\'e}nic post-shock region, magnetic field lines tend to have a weaker undulation than the dense sub-layer, which leads to a difference in the angle between the sub-layer and the surrounding post-shock magnetic fields (see Figure~\ref{vxconv}) at different locations along the sub-layer. In regions where the angle is larger, more material is accumulated onto the sheet through the nearly field-aligned secondary converging flows, which can condense further gravitationally. In this picture, they are the birth sites of overdense filaments and cores. 
%While the initial corrugation generally follows the spatial variation of the inflow velocity, its shape and relative orientation with respect to the surrounding magnetic field are crucial in determining the evolution of local gas content. Corrugation along the magnetic field with a distinct angle between the thin layer and the field line becomes a ``preferred location" where the secondary convergent flow is the most efficient, and develops into the birth site of overdense filament or cores. Corrugations almost parallel to the magnetic field, on the other hand, appear to be no-collision points with no significant density enhancement.

When the perturbative inflow velocity field varies perpendicular (rather than parallel) to the post-shock magnetic fields (i.e.~with a spatial dependence in the $y$-direction in our model), the situation is quite different.  This is because the perturbation tends to roll up the dense sub-layer around an axis that is nearly parallel to the dominant component of the magnetic field threading the sub-layer, which leads to little field bending and is thus not resisted by magnetic tension. This lack of magnetic resistance makes the rolling of the dense sheet relatively easy, producing corrugations of the sub-layer that, when viewed from a direction other than edge-on, appear as magnetically aligned striations with higher column densities than their surroundings that are prevalent in both our simulations and apparently in interstellar clouds as well. The contrast in column density comes principally from the difference in the path length through the (corrugated) sub-layer at different locations, rather than any variation in volume density. As a result, the apparent column density enhancement is usually relatively moderate (within a factor of a few). This is very different from the case of field-misaligned massive filaments, which can grow to a much higher volume density than their surroundings through mass accumulation along the magnetic field.  
%Ridges perpendicular to the post-shock magnetic field evolve differently. Since the magnetic pressure is suppressing gas flows across the magnetic field lines, materials are prohibited to move between corrugations. Instead, the shearing force from imbalance between upstream ram pressure and post-shock thermal pressure, if strong enough, can roll up the thin sub-layer and twist the magnetic field altogether without bending the field lines. The folding of the sub-layer results in longer path length when integrated along the inflow direction, which corresponds to denser structure in the column density map. We therefore see thin, horizontal (parallel to the post-shock magnetic field) stripes in the column density when there are variations in the inflow velocity perpendicular to the post-shock magnetic field.

\subsection{Prestellar Cores and Striations in Turbulent Cloud Simulations} 
\label{sec::dis::core}

\begin{figure}
\begin{center}
\includegraphics[width=\columnwidth]{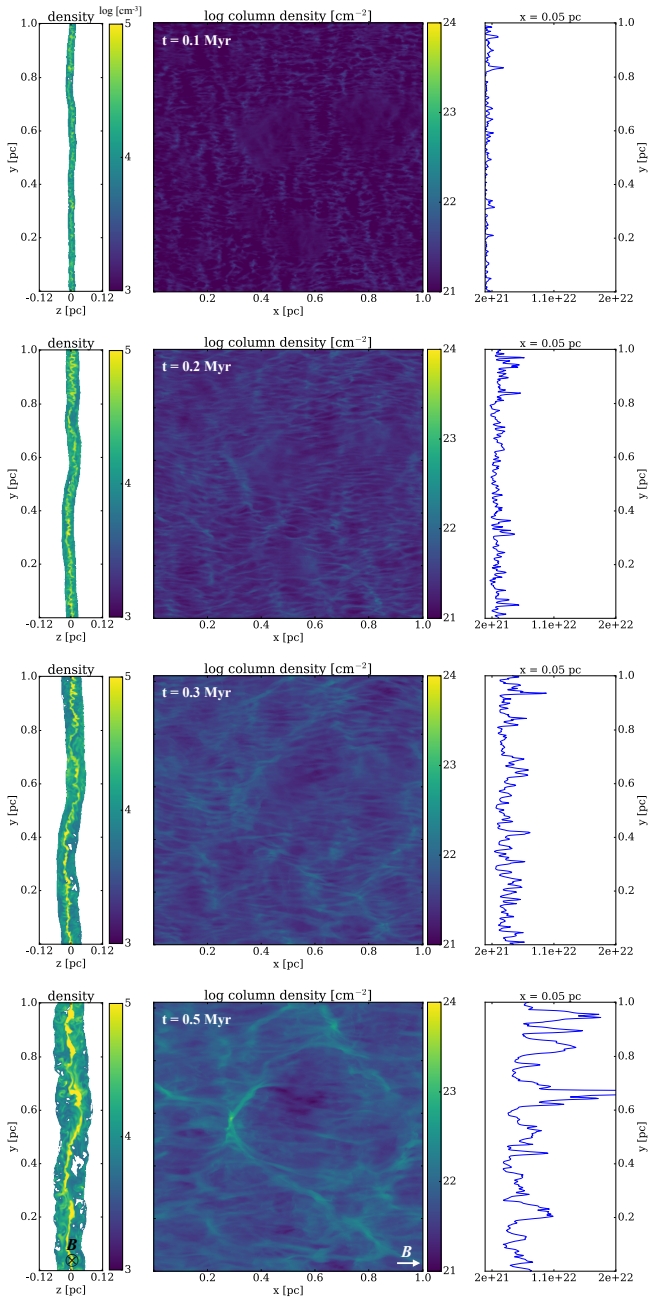}
\caption{Evolution of the post-shock layer formed in the core-forming simulation discussed in CO15 (model M10B10). {\it Middle panel:} column density map integrated along the inflow direction. {\it Left and right panels:} density structure ({\it left}, in log scale) and column density line plot ({\it right}) of a slice cut through $x=0.05$~pc. This location is chosen to include some visible striations while avoiding any prominent filaments and cores. Corrugation of the central sub-layer can be clearly seen from the density map, and the evolution of the column density line plot is very similar to Figure~\ref{sinvy} that small perturbations grow over time to become visible gas structures. }
\label{evoM10}
\end{center}
\end{figure}

We extended the same analysis to the original core-forming simulation described in \hyperlink{CO15}{CO15} (Figure~\ref{M10map}) to see whether the same explanation is applicable when multi-scale turbulence is present. Figure~\ref{evoM10} illustrates the evolution of the core-forming simulation, showing the column density map (integrated along the large-scale inflow, middle panel) and the line plot of column density along $y$ at $x=0.05$~pc (right panel), which cut through many striations but no overdense cores/filaments. The density map of the $y$-$z$ slice is also shown in the left panel of Figure~\ref{evoM10}. Both the corrugation of the sub-layer in the density map and the growth of the perturbation in the column density line plot are very similar to the simpler cases discussed in previous sections (see Figures~\ref{sinvy} and \ref{sinvwx}). The similarity supports the notion that the striations in these more turbulent clouds are also corrugations of the stagnated dense sub-layer in the center of the post-shock region integrated along the line of sight. 

\subsection{Connection with Observations}

Our simulated striations formed in the corrugated sub-layer are in good agreement with the observed ones in that they both align with ambient magnetic field. 
In addition, the CO molecular line observations towards the Taurus molecular cloud report clear presence of striations in the lower-level emission of $^{12}$CO, and striations are relatively faint in $^{13}$CO emissions \citep{2008ApJ...680..428G,2016MNRAS.461.3918H}. This implies that the volume density of striations is not significantly enhanced from the diffuse gas; therefore the higher column density within striations (so they could be distinguished from the background) can only be explained by integrating through longer path lengths, which is the case of the corrugated sub-layer model.

Dynamically, since the sub-layer is stagnated from the secondary convergent flow, striations formed within it are also quiescent structures. This is consistent with the striations appearing in velocity channel maps of both CNM and MCs \citep{2006ApJ...652.1339M,2016MNRAS.461.3918H}, in that there is almost no velocity variation along individual striations. On the other hand, the fact that striations are visible across several velocity channels in a local area ($\sim 2-4$~km/s) can be explained by different centroid velocities of multiple corrugations moving towards or away from the observer along the line of sight (see e.g.~the left panel of Figure~\ref{sinvy}).

%G+08: in the lower level emission / velocity 5-8 km/s / subthermally excited / N ~ 2e21 cm^-2 (~2x background; relatively diffuse) / parallel to B measured by optical starlight polarization / **spatially coherent within velocity intervals less than 0.25 km/s** (H+16)
%H+08: velocity 5.5-7.5 km/s
%H+16: channel images of 12CO and 13CO / evident within the core velocity interval 6.3-6.7 km/s / also have faint features within 7.38-7.72 km/s / Figure 1: no significant velocity variation within single striation! / quasi-periodic pattern/velocity oscillation / faint 13CO compared to 12CO

Moreover, the nearly fixed spacing between striations (see e.g.~the spectrograms in \citealt{2016MNRAS.461.3918H}) is likely the result of a dominant scale of turbulent velocity in the direction perpendicular to the magnetic field. As discussed in Section~\ref{sec::vy}, NTSI is most efficient at certain locations, which are determined by the spatial variation of the large-scale velocity (see Figure~\ref{sinvwx}). The spacing between striations might therefore reflect the spatial frequency of the cloud-scale turbulence, though more detailed examinations are needed to quantitatively establish this connection.

\section{Summary}
\label{sec::sum}

In this study, we explored the shock-compressed regions under conditions similar to those within interstellar clouds. Analytical solutions describing the post-shock gas dynamics are provided. We identified and characterized a dense, stagnated sub-layer in the post-shock region induced by a secondary convergent flow, which is a unique feature of strong oblique MHD shocks; it disappears in the absence of a dynamically significant magnetic field. This thin sub-layer, when corrugated, is subject to NTSI. Because the sub-layer is naturally threaded by a strong magnetic field that lies nearly in the plane of the layer, it is relatively easy to roll around the field lines when perturbed, resulting in those striations parallel to the local magnetic field in the integrated column density map, as commonly seen in observations and simulations.

The major conclusions of this study are summarized as follows:
\begin{enumerate}
\item The oblique secondary flow in the post-shock region (Figure~\ref{sublayer}) is an important feature of oblique MHD shocks (Equation~(\ref{thetav})). When strong enough (Equation~(\ref{v'})), these secondary flows become shocks and further compress gas in the post-shock region to form a thin, dense sub-layer (Figures~\ref{sublayer} and \ref{sublayer_sim}), which provides the preferred environment (high density, relatively low magnetization) for overdense structures like prestellar cores to form (see e.g.~Figure~\ref{M10map}).
Roughly speaking, the sub-layer is formed when the pre-shock Alfv{\'e}n speed is larger than the sound speed. It should be prevalent in interstellar clouds since its formation conditions are relatively easy to satisfy.  

\item Because of the secondary convergent flow, this dense sub-layer is stagnated when created (Figure~\ref{vxcorr}). 
This supports the anisotropic core formation model discussed in \hyperlink{CO14}{CO14} and \hyperlink{CO15}{CO15} that gas movement along the magnetic field direction near prestellar cores is mostly induced by core's gravity. 
The striations formed out of the sub-layer are also stagnated features, meaning that they are not feeding or moving towards nearby massive filaments (Figures~\ref{sam1} and \ref{sam2}), which agrees with the observations that there is no significant velocity gradient along striations \citep[see e.g.][]{2016MNRAS.461.3918H}. 

\item When the sub-layer is perturbed by the cloud-scale turbulence or any spatial variation of the gas velocity in the pre-shock region, the shape and orientation of the rippled sub-layer are key to the formation and evolution of any overdense structure within the sub-layer. 
This is especially true for the angle between the sub-layer and the post-shock magnetic field when the perturbation is spatially varying along the magnetic field 
%Ripples parallel to the magnetic field controls the locations where further condensation can take place through the angle between the sub-layer and the local magnetic field 
(Figures~\ref{sinvx} and~\ref{vxconv}). 
Since the post-shock gas flow roughly follows the direction of the magnetic field, locations with larger angles between the sub-layer and the magnetic field would result in more efficient secondary convergent flow and stronger compression, and become seeds of overdense clumps.

\item Because of a lack of resistance from the magnetic tension force, perturbations in the inflow that vary spatially perpendicular to the post-shock magnetic fields can easily roll up the thin sub-layer around the major field direction, making the dense sheet highly corrugated in the direction perpendicular to the dominant component of the magnetic field but not along it. The corrugation is further enhanced by the (MHD) 
%the magnetic tension, there is no material flow along ripples perpendicular to the magnetic field. However, the shearing force from the imbalance between ram pressure of the large-scale turbulence and the thermal pressure of the dense sub-layer could in principle be strong enough to fold and roll up the thin sub-layer and local magnetic field (Figure~\ref{vyLvy}). The sub-layer is therefore further bent under the MHD 
non-linear thin-shell instability (NTSI), and thin, elongated stripes parallel to the magnetic field emerge in the column density map from integrating over longer path lengths within the corrugated sub-layer (Figures~\ref{sinvy} and \ref{sinLvy}).

\item Our picture of the formation of diffuse striations and massive filaments is reinforced by idealized simulations with inflow velocity perturbations that vary spatially both along and perpendicular to the post-shock magnetic field
%Simulations with turbulence with spatial variations in both direction (parallel and perpendicular to the post-shock magnetic field) show combined effect from corrugated sub-layer 
(Figure~\ref{sinvwx}). Striations parallel to local magnetic field are still built up from rolling/bending the sub-layer around field lines, but with reduced lengths controlled by the relative orientation of the sub-layer locally with respect to the magnetic field, which defines ``preferred locations" where condensed clumps would appear. The idealized simulations capture the essence of the gas structures formed in more realistic simulations with multi-scale turbulence (Figure~\ref{evoM10}).

\end{enumerate}

To conclude, the success of the corrugation of dense sub-layer for explaining the origin of striations is very encouraging and provides strong motivation for testing these ideas in observations. Further studies considering more observable parameters like polarization fraction and velocity dispersion within striations at various viewing angles would also be interesting to examine the properties of the stagnated post-shock sub-layer and explore potential tracers of the direction and strength of local magnetic field in the presence of prominent striations.

\acknowledgements
C.-Y.~C. thanks the support from Virginia Institute of Theoretical Astronomy (VITA) at the University of Virginia through the VITA Postdoctoral Prize Fellowship. 
Z.-Y.~L. is supported in part by NSF AST1313083 and NASA NNX14AB38G. 
P. K.~K. is supported by the National Radio Astronomy Observatory (NRAO) through a Student Observing Support Grant and by the Jefferson Scholars Foundation and the Virginia Space Grant Consortium through graduate fellowships. 
L. M.~F. is a Jansky Fellow of NRAO. NRAO is a facility of the National Science Foundation (NSF operated under cooperative agreement by Associated Universities, Inc).

\end{document}